\documentclass[journal]{IEEEtran}

\usepackage{cite}
\usepackage{authblk}
\usepackage{xcolor}
\usepackage[draft]{todonotes}
\usepackage{nth}
\usepackage{amsmath}
\usepackage{booktabs}
\usepackage[ruled]{algorithm2e}
\usepackage{tabularx}
\usepackage{abstract}

\usepackage{microtype}

\title{TG-PSM: Tunable Greedy Packet Sequence Morphing Based on Trace Clustering}

\author[1]{Farzam Fanitabasi}
\affil[1]{Department of Computer Engineering, Sharif University of Technology.
\{ffani,amakbari\}@ce.sharif.edu}

\begin{document}
\twocolumn[
\begin{@twocolumnfalse}
\maketitle
\end{@twocolumnfalse}

\setlength{\absleftindent}{0pt}
\setlength{\absrightindent}{0pt}

\begin{abstract}
\noindent
Common privacy enhancing technologies fail to effectively hide certain statistical aspects of encrypted traffic, namely individual packets length, packets direction and, packets timing. Recent researches have shown that using such attributes, an adversary is able to extract various information from the encrypted traffic such as the visited website and used protocol. Such attacks are called traffic analysis. Proposed countermeasures attempt to change the distribution of such features. however, either they fail to effectively reduce attacker's accuracy or do so while enforcing high bandwidth overhead and timing delay. In this paper, through the use of a predefined set of clustered traces of websites and a greedy packet morphing algorithm, we introduce a website fingerprinting countermeasure called TG-PSM. Firstly, this method clusters websites based on their behavior in different phases of loading. Secondly, it finds a suitable target site for any visiting website based on user indicated importance degree; thus providing dynamic tunability. Thirdly, this method morphs the given website to the target website using a greedy algorithm considering the distance and the resulted overhead. Our evaluations show that TG-PSM outperforms previous countermeasures regarding attacker accuracy reduction and enforced bandwidth, e.g., reducing bandwidth overhead over 40\% while maintaining attacker's accuracy.
\end{abstract}

\begin{IEEEkeywords}
Traffic Analysis, Website Fingerprinting Countermeasure, Encrypted Traffic Morphing,Website Clustering
\end{IEEEkeywords}

\bigskip]

\section{Introduction}
Website fingerprinting (WF) attacks leverage the shortcomings of common privacy enhancing technologies (PET) such as SSH, SSL and Tor in order to infer the visited website of a monitored user (MU). Such PETs employ different encryption techniques in order to preserve the privacy of users. However, they fail to effectively hide or obfuscate features such as packets length, packets direction and packets timing. Using these coarse features, an attacker is able to  perform WF attacks and gain insight into website, which user is visiting or has already visited. \par

The mentioned PETs shortcomings are due to the fact that WF attacks inspect network traffic in a different abstraction level compared to usual PETs. Hence, they are able to reveal some information that common PETs are expected to hide in the first place. This difference in abstraction level inspection is caused by the PETs failure to hide at least one of the following traffic feature classes: packets length, packets direction, packets timing ,and packets sequence. Furthermore, each feature class can be represented in different granularities, forming a hierarchical order, e.g. packets form flows, flows form objects, objects form traces ,and so on. WF attacks use such features in different abstraction and granularity levels to model an interesting application behavior.  \par

Although WF attacks have a fairly simple threat model, recent researches\cite{dyer2012peek,bissias2006privacy,liberatore2006inferring,cai2014systematic,panchenko2011website} have shown that they are capable of achieving high accuracy rates and pose serious threats to online privacy of the users. Even with the presence of a PET and a WF countermeasure, in 2012, Dyer et.al. \cite{dyer2012peek} showed that all such defenses either fail to reduce the accuracy of the attacker to a satisfying measure or do so with high bandwidth and/or timing overhead. For example, Dyer developed a simple na\"{\i}ve Bayes classifier  trained with byte count in traffic bursts as the sole feature. This classifier was able to detect the visited website of a user with lower bound accuracy of 46\% out of 128 websites. \par

The proposed defenses in this field, either change the distribution of at least one class of traffic features \cite{Sun02} such as random padding and exponential padding, or create uniform traffic traces, aiming to limit the distinguishability between different traffic instances, such as BuFLO \cite{dyer2012peek} and TAMARAW \cite{cai2014systematic}. In addition, another type of defense belonging to the first approach are morphing defenses \cite{WrightMorphing09}. Morphing defenses change the distribution of features in a way that the observed trace resembles another website. Defenses pursuing the first approach use the means of padding and splitting in attempt to obfuscate the real packets length of the loaded site, or utilize application level features such as HTTP pipelining in HTTPOS \cite{HTTPOS}. The main difference between morphing defenses and the other ones, is that morphing defenses mislead the attacker while the others confuse the attacker resulting in drop-in-accuracy percentage (DiP). The second approach on the other hand, limits the information leakage by uniformization, thus removing the distinctive features among website traffic instances. Such defenses are inspired by MixNets \cite{danezis2003mixminion} and use uniform size packets at predetermined and fix time intervals. Although this approach provides better DiP than the previous defenses, they suffer from high bandwidth and/or timing overhead \cite{dyer2012peek}. High bandwidth overhead is problematic in bandwidth-starved applications such as Tor and defenses with excessive timing delays are not suitable for time-sensitive applications such as chatting and mailing. \par

In addition, none of these defenses support dynamic tunability. This means their relative DiP and overhead is tuned once at the design phase. Moreover, changing these attributes depending on the user input and/or the importance of the given instance is not supported. This ability is particularly important in scenarios where all websites are not equally important to users. In such cases, the user might prefer to avoid high overhead for a website, which is not important to its privacy and vice-versa. In these scenarios, dynamic tunability can help the user decide how important a particular website is to his/her privacy and then tune the defense accordingly.  \par

In this paper, we introduce a novel morph-based countermeasure against WF attacks called TG-PSM. Unlike the previous morphing defenses that only morph the packet size distribution of the source site, TG-PSM analyzes the packet sequence itself and morphs the size and timing of each packet according to its greedy packet morphing function. TG-PSM has two main components: (a) greedy morphing and (b) website clustering. The greedy morphing component is in charge of morphing the source website trace to the target website trace. This morphing is done in a greedy manner, i.e., for morphing each packet, the subsystem maximizes the goal function. This function compares the relative DiP and enforced overhead of morphing that particular packet. The second component, website clustering, is in charge of choosing the appropriate target site. For each given website that user wants to hide, an importance measure (IM) is required. Based on the given website and its IM, the subsystem chooses a cluster of websites and extracts a random website from that cluster as the target site. This difference of target site selection based on user input provides TG-PSM with dynamic tunability. The contributions of this paper are as follows:
\begin{table}[t]
\centering\small
\begin{tabular*}{\linewidth}{@{\extracolsep{\fill}}p{0.3\linewidth}p{0.9\linewidth}@{}}
\toprule
Symbol & Meaning  \\
\midrule
	 WF & Website Fingerprinting \\
         PET & Privacy Enhancing Technologies \\
         MU & Monitored User \\
         DiP & Drop in Accuracy Percentage \\
         WS & Set of All Websites \\
         TR & Set of All Traces \\
         TF & Set of All Traces with Their Feature Space \\
         Cand & Set of All Candid Traces \\
         IM & Importance Measure \\
         nClust & Number of Website Clusters \\
         $T_{o}$ & Initially Observed Trace by Attacker \\
         $T_{src}$ & Source Website \\
         $C_{src}$ & Cluster of Source Website \\
         $C_{dst}$ & Cluster of Destination Website \\
         $T_{dst}$ & Destination Website \\
         $T_{out}$ & Output Trace with Countermeasure Applied \\
         $CD$ & Cluster Distance Matrix \\
         $CD[i,j]$ & $jth$ farthest cluster from cluster $i$ \\
         M & Number of Slices for Trace Representation \\
         SimThres & Similarity Threshold \\
         \bottomrule
\end{tabular*}
\newline
\caption{Used Symbols in this Paper}
\label{UsedSymbols}
\end{table}

\begin{itemize}
  \item A  greedy algorithm based on sequence morphing, which provides packet morphing without requiring early distribution of packets.
  \item A  technique for presenting websites based on their behavior in different phases of their traffic as well as clustering them based on known cluster methods. Also, we show the performance of each clustering method using internal and stability measures and discuss the best clustering algorithm for trace clustering.
  \item A dynamically tunable defense against WF attack is introduced. This tunability is based on choosing the appropriate cluster for target site based on the user specified IM.
\end{itemize}
\par
To evaluate our method (TG-PSM), we implemented 3 classifiers and 3 countermeasures from the literature and compared them to TG-PSM based on their DiP and bandwidth overhead. The classifiers include Liberatore and Levine's na\"{\i}ve-Bayes \cite{liberatore2006inferring}, Herrmann's multinomial na\"{\i}ve-Bayes \cite{herrmann2009website} and Panchenko's SVM classifier \cite{panchenko2011website}. For countermeasures, we chose the ones that have already shown effective DiP against leading classifiers or related to TG-PSM family of countermeasures. The countermeasures are Traffic Morphing, BuFLO and TAMARAW. Our results show that in same scenarios, TG-PSM performs better than the rest of the countermeasures. It reducing bandwidth overhead by more than 40\% while it maintains the same DiP. Used symbols and abbreviation in this paper are presented in Table \ref{UsedSymbols}

The rest of this paper is organized as follows. Section II and III cover preliminary concepts in WF field by reviewing website fingerprinting attacks and feature class categorization, respectively. In section IV, the related work of this field both in attacks and defenses are presented. Section V, explains the morphing algorithm and the process of finding the suitable target website. Section VI, presents the evaluation results of TG-PSM and compares the effectiveness and enforced bandwidth of our method to other well known defenses against best-performing attacks of this field. Section VII discusses some important subjects in WF attacks and defenses as well as practical considerations. Section VIII conveys our discussion and provides concluding remarks. 

\section{Website Fingerprinting Attacks}

WF attacks are considered as a subset of traffic analysis attacks. In WF attacks, the threat model consists of a passive attacker with the means to relate the observed traffic to a user (MU). Attacker also has the ability to sniff the ongoing traffic of the network. Because of the simplicity of the threat model and the lightweight requirement for the attacker, WF attacks are virtually undetectable, hence they can be categorized into untraceable surveillance methods \cite{herrmann2015online}. An attacker might acquire the means to relate an observed traffic to a user if it possesses one of the three following conditions: (a) the attacker is local, (b) the monitored channel is used only by a single user and (c) one of the nodes between client and server, which knows the identity of the client, is compromised \cite{Ling09,houmansadr2013need}. This situation is possible in proxy servers with authentication or entry node of a Tor circuit. \par
WF attacks consist of three steps \cite{nguyen2008survey}. In the first step, the attacker specifies a set of interesting websites (WS). This set includes the websites that the attacker is interested in knowing if they are being visited. Then, the attacker browses such websites in a similar environment close to the users. While browsing, the attacker captures the data and stores it for further use ($T_{o}$). In second step, the attacker creates a statistical model based on the $T_{o}$, which describes the traffic behavior while the website is being visited. Here, the attacker might use various learning algorithms or a conjunction of multiple algorithms in order to create the model \cite{nguyen2008survey}. In the final step, the attacker sniffs the ongoing traffic and matches the ongoing traffic pattern with the patterns of the interesting websites. For any given traffic trace, if the similarity of the trace to an interesting website passes a certain threshold, attacker infers that MU is visiting that particular website.\par
Consistent with previous researches, we consider the following assumption in our framework:
\begin{itemize}
\item
Attacker knows when a trace for a single website starts and when it is ended. In the real world, this assumption holds if MU injects a delay in between visiting of two different websites.
\item
No other conflicting background activity exists at the time of sniffing. This background activity should not interfere with the website visit in the way that it confuses the attacker, for example, into detecting that user is downloading a file instead of loading a webpage. In general, even multi tab browsing can be considered a conflicting background activity. The result of multi-tab browsing and some other important factors are discussed in \cite{Juarez2014}.
\end{itemize}

\section{Important Traffic Features}

In order for the classifier distinguish between different instances of website traffic, classifier needs a statistical model, which accurately describes each websites behavior. As creating a model based on raw traffic (only storing packet information) have proven to be difficult and inaccurate, different classifiers extract various features out of the traffic. Such features can be categorized into these four classes: (a) unique packet length, (b) inter packet timing, (c) packet sequence behavior and (d) packet direction. \\
We define a packet as an ordered triplet of direction (-1 for outgoing and 1 for incoming), $t_{i}$ as the absolute time difference of this packet with the one before and size:
\begin{gather*}
	Packet\: p_{i}\:=\:<dir_{i},t_{i},size_{i}>  \\ 
	\:dir_{i} \in \{-1,1\} \: and  \:t_{i}  = (time_{p_{i}} - time_{p_{i-1}} )
\end{gather*}
	A packet sequence as an ordered set of $n$ packets each with their own attribute:
\begin{gather*}
	Packet Sequence\:PS\:=\: <p_{1},p_{2},\ldots,p_{n}> 
\end{gather*}
	A trace as an ordered pair of $site_{id_{j}}$ of which the trace belongs to, and the packet sequence for that trace:
\begin{gather*}
		Trace\:t_{j}=<site_{id_{j}},ps_{j}> \\
		site_{id_{j}} \in WS \: and \: PS_{j} \in PSs 
\end{gather*}
A website as another ordered pair of $site_{id_{k}}$ and $Tr_{j}$ as the set of traces belonging to a specific website: 
\begin{gather*}	
	Website\:ws_{k}\:=\:<site_{id_{k}},Tr_{j}> \\
	Tr_{j} \subset TR
\end{gather*}	
Based on this description, now we can define the four important feature classes in WF attacks. Two traces $t_{i}$, $t_{j}$ are different regarding to their first class features if:
\begin{gather*}
t_{i}: \: <site_{id_{i}},ps_{i}> \:,\: t_{j}: \: <site_{id_{j}},ps_{j}> \\
(\exists p_{r} \in ps_{i} \: | \: p_{r} \not\in ps_{j}) \: \vee \: (\exists p_{r} \in ps_{j} \: | \: p_{r} \not\in ps_{i})
\end{gather*}
Two packets of the same size in two different traces have different arrival time (different regarding to the second class of features) if:
\begin{gather*}
\exists \: p_{p} \in ps_{i} \:\land\: p_{q} \in ps_{j} \: and \: size(p_{p})=size(p_{q}) \\
SUM(t_{h}) \neq SUM(t_{u}) \: where \\
t_{h} = \{t_{p_{h}} \: | \: 1 \le h \le p\} \: and \\
t_{u} = \{t_{p_{u}} \: | \: 1 \le u \le 1\} 
\end{gather*}
Two packet sequence show different behavior regarding to the same metric SM if:
\begin{gather*}
SM(ps_{i}) \neq SM(ps_{j})
\end{gather*}
For example if the SM represents the mean packet size across first 50 packets of each trace:
\begin{gather*}
MEAN(p_{x}) \neq MEAN(p_{z}) \: where \\
p_{x} = \{p_{x} \in ps_{i} | 1 \le x \le 50\} \: and \\
p_{z} = \{p_{z} \in ps_{j} | 1 \le z \le 50\} 
\end{gather*}
The use of packet sequence behavior features, can decrease the dependency of attacks to the packet specific features thus increasing the overall accuracy of classifiers in face of noise and defense. Panchenko et. al. \cite{panchenko2011website} used various packet sequence features in their attack (they called these features markers).

\section{Related Work}

As the literature of this field is vast and diverse, we focus on the related work which directly affect our approach or have made significant improvements in this field. 
\subsection{WF Attacks}

We categorize WF attacks based on their approach to similarity measures and the type of learning algorithm they used\footnote {In \cite{wang2013improved} Wang et.al. has made a similar categorization, however, based on different factors}. Using these two attributes, two main WF attack categories are: feature attacks and distance-based attacks. \par

\subsubsection{Feature Attacks}

Feature attacks at best utilize only 3 of the feature classes and leave out the packet sequence behavior class. They use simple learning algorithm such as na\"{\i}ve-Bayes or Jaccard-coefficient \cite{han2011data}. Early researches in this area dates back to early 2000s when Sun et.al. \cite{Sun02} analyzed log files of web traffics and identified different websites based on their object size. Though at the time this classifier performed well, new modifications in HTTP protocol such as HTTP pipelining \cite{lu2010website} and browser design such as multiple concurrent connections have made this attack insufficient. In 2006 Liberatore and Levine \cite{liberatore2006inferring} introduced two new classifiers. A Jaccard coefficient and a na\"{\i}ve Bayes classifier. Both of these classifiers focus on unique packet length and ignored other feature classes. Both worked in closed world environment\footnote{A discussion about close and open word environment is presented in section VII} and only considered the traffics which went through a SSH proxy. In 2009 Herrmann et.al. \cite{herrmann2009website} improved the work of Liberatore and Levine and introduced a new classifier that used multinomial na\"{\i}ve Bayes algorithm and in addition to the unique packet lengths, also considers the frequency of each packet. Herrmann classifier was tested against various PETs at the time and with the exception of Tor, it performed reasonably well. All of these attacks suffer from their extensive focus on unique packet length feature class and fail to comply with defenses that change these features. This shortcoming is more apparent in situations with high noise such as Tor. However, for comparability with the literature, we use Liberatore and Levine (LL) classifier and Herrmann (HA) classifier in our evaluations. \par

\subsubsection{Distance-Based Attacks}

Distance-based attacks usually exploit either of these two main approaches. They either use distance-based learning algorithms such as SVMs or string distance measures such as OSAD and considered traffic traces as strings. The first approach relies mostly on packet sequence behavior class. The second approach on the other hand, doesn't extract features out of the traffic, instead creates a string out of the observed packets. Most of attacks in this class focus on attacking Tor streams because of their uniform packet lengths. In 2011 Panchenko et.al. \cite{panchenko2011website} introduced an SVM classifier, which in addition of using packet length and timing features, used various packet sequence metrics (which they called markers) to better classify instances. This classifier (PA) was one of the first to operate reasonably well in Tor environments. In 2010, Lu et.al. \cite{lu2010website} used levenshtein string similarity measure and build a classifier specifically for Tor. Use of string similarity metrics was also pursued in \cite{wang2013improved,cai2014systematic,Cai2012} and some improvements was done on their speed and scalability.  In 2012 Cai et.al. \cite{Cai2012} introduced DLSVM based on Damerau-Levenshtein distance and uses an SVM learning algorithm. Although  they showed that DLSVM performs better the PA on Tor traffic, however, in the case of normal traffic, PA performs marginally better. In 2013 Wang and Goldberg \cite{wang2013improved} improved on previous Cai et.al. classifier and introduced OSAD (optimal string alignment distance). Even though OSAD showed high accuracy even in Tor situations, however, its excessive training time (608.000 CPU second for 100 websites 40 instances each) is a major disadvantage. In 2014 Wang et.al. \cite{wang2014effective} introduced $k$-NN classifier, which has 6 different feature classes and relies on large datasets to find feature classes that defenses are unable to hide. 

\subsection{WF Defenses}

We study WF defenses based on six different evaluation metrics. The first metric shows the OSI layer in which the defense works. Defenses that only operate on network level are restricted in using only network level utilities. They are able to perform network-level modifications to traffic and are application independent. On the other hand, layer 7 defenses have the capability to use application level modifications that in some cases increases their DiP. Another metric refers to the feature class(es) each defense is attempting to hide (OF(s)). Intuitively, the more feature classes defense hides, the higher DiP it has. The trade-off here is that hiding all of the classes results in high overhead that is undesirable. Third metric is determinism, which shows the predictability of $T_{out}$. If for any reason the attacker learns the used defenses mechanism, non-deterministic defenses can still operate while deterministic defenses become predictable. Forth metric is tunability. This metric shows the ability of a defense mechanism to tune its relative overhead and DiP. If a defense is tuned once and uses the same approach to all instances it has static tunability. However if the defense mechanism can change its attributes for each instance based on users input, it has dynamic tunability. This feature is specially important in the case of bandwidth-starved situations. In these scenarios, the user does not need to tolerate high overhead for all its actions and can chose to afford more overhead only for more important cases. The last two metrics are bandwidth overhead and timing delay that both are important in designing a useful and efficient defense.\par
In \cite{hintz2003fingerprinting} Hintz et.al. proposed numerous countermeasures such as random padding and pad to MTU, however, they showed that those countermeasures were ineffective. In 2009 Wright et.al. \cite{WrightMorphing09} proposed direct target sampling (DTS) and traffic morphing (TM). Both of these defenses morphed the shape of $T_{src}$ and made it resemble another instance. The difference was, while DTS used random target selection and morphed the $T_{src}$, traffic morphing used convex optimization matrix and morphed to $T_{dst}$ with low bandwidth overhead.
While all previous countermeasures only used network level utilities, in 2010 and 2011, Panchenko \cite{panchenko2011website} and Luo \cite{HTTPOS} introduced Camouflage and HTTPOS, respectively. Camouflage worked by requesting multiple other websites while the actual website was visited by the user to cloud the real traffic. HTTPOS on the other hand used conjunction of application layer techniques such as HTTP pipelining and network level such as TCP RTT to reform the $T_{out}$. The most recent defenses such as BuFLO and TAMARAW are inspired by high latency MixNets. These countermeasures attempt to hide most distinctive features of a traffic by sending fixed length packets and fixed intervals. The only difference between TAMARAW and BuFLO is that the former one performs more efficiently as it tunes packets interval time based on volume of incoming and outgoing traffic, i.e., giving higher packet rate to incoming and less to outgoing flows. Although these countermeasures enjoy a very good DiP, however, they enforce high overhead and timing delay on network and are (as mentioned) unsuitable for bandwidth starved scenarios such as Tor. Another more recent counter measure is Glove by \cite{nithyanand2014glove}. Glove considers traces as time series data and clusters them based on DTW (dynamic time warping). Then for each cluster, it creates a super trace. For any given cluster, Glove morphs the trace into its clusters super trace and then sends it. The comparison between different defenses can be seen in Table \ref{DefenseComparison}.

\begin{table*}[ht]
\centering
\begin{tabular}{p{0.13\linewidth}p{0.1\linewidth}p{0.1\linewidth}p{0.1\linewidth}p{0.1\linewidth}p{0.1\linewidth}p{0.1\linewidth}}
\hline
Defense  & OSI Layer(s) & OF(s) & Determinism & Tunability &BWO & TD \\
\hline
PadR & 3 & PL & No & No & Low & None\\
PadMTU & 3 & PL & Yes & No & No & None\\
Mice/Elephant & 3 & PL & No & Static & Low & None\\
Traffic Morphing & 3,7 & PL & No & Static & Medium & None\\
Camouflage & 7 & PL,Ti & No & Static & TSD & None\\
HTTPOS & 3,7 & PL,Ti & No & Static & Medium & Medium\\
BuFLO & 3 & PL,Ti,PS & Packet Level & Static & Very High & Very High\\
TAMARAW & 3 & PL,Ti,PS & Packet Level & Static (2Way) & High & High\\
Glove & 3,7 & PL,Ti & Cluster Level & Static & Low & None\\
\textbf{TG-PSM} & 3,7 & PL,Ti,PS & No & Dynamic & TSD & Medium\\
\hline
\end{tabular} 
\newline
\caption{Comparison Between WF Defenses.OF(s) refers to the features that defense tries to obfuscate. BWO is bandwidth overhead and TD is timing delay. TSD refers to target site dependent.}
\label{DefenseComparison}
\end{table*}

\section{Proposed Method: TG-PSM}

\begin{figure} [h]
\centering
\includegraphics[width=9cm, height=5cm]{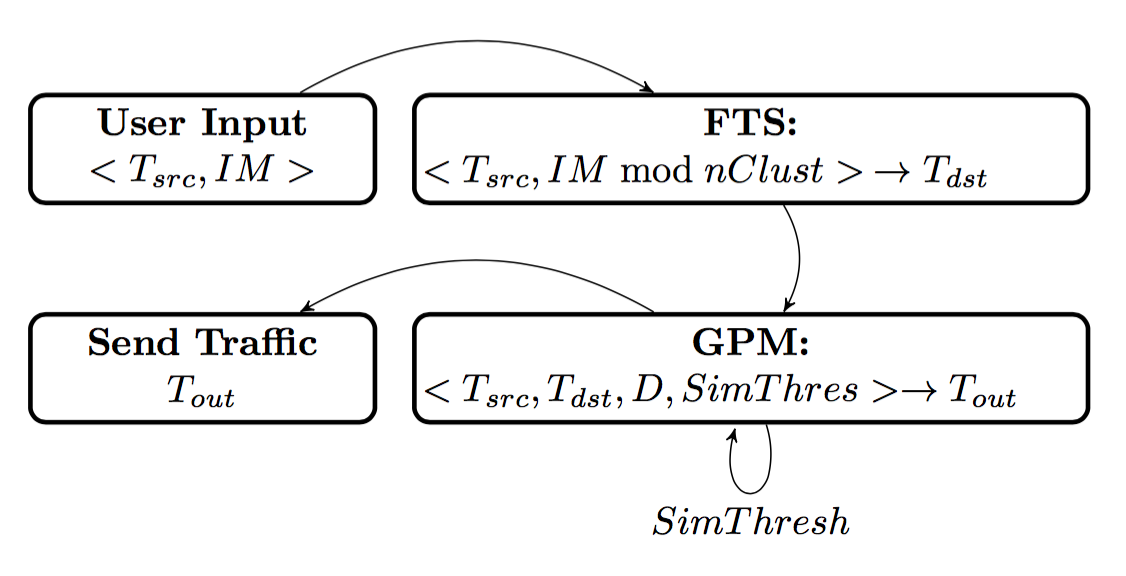}
\caption{Overall components and respective steps in TG-PSM}
\label{TG-PSM}
\end{figure}

In this section, we introduce TG-PSM. The overall components of TG-PSM is shown in Figure \ref{TG-PSM}

\subsection{Trace Morphing}

In TG-PSM, trace morphing is the step in which the real traffic trace generated by a user ($T_{src}$) is changed to look like another trace ($T_{dst}$). Trace morphing is performed by changing the size and timing of each packet of $T_{src}$. Morphing is done by using TCP protocol features such as packet padding, packet fragmentation, and packet delaying. We define source trace ($T_{src}$) as the traffic trace generated by MU, destination trace ($T_{dst}$) as the traffic trace chosen for $T_{src}$ to morph into, and $T_{out}$ as the trace produced by applying TG-PSM algorithm to $T_{src}$:
\begin{equation}
 T_{out} = TG-PSM(T_{src},T_{dst},D,SimThres).
\end{equation}

Source trace is the result of user activities such as browsing the web or downloading a file. $T_{src}$ is not available at once to the morphing algorithm. As a result, TG-PSM is designed as a stream processing algorithm over traffic packets that may only hold and delay a small number of packets at any time during the morphing. Morphing is done on MU's device such that any adversary monitoring the traffic will only see $T_{out}$. For each traffic flow, $T_{dst}$ is selected based on $T_{src}$ and the IM specified by MU, such that detecting generated traffic as $T_{dst}$ has the least privacy implications for MU. Details of selecting $T_{dst}$ based on $T_{src}$ and the tuning parameters (IM) set by the user is discussed in section B. Morphing $T_{src}$ to $T_{dst}$ is done considering overhead of morphing and similarity of resulting trace ($T_{out}$) to $T_{dst}$. Therefore, $T_{out}$ may not completely resemble $T_{dst}$ in order to maintain some constraints on bandwidth overhead and timing delay. The goal function and similarity measures used in determining such trade-offs are discussed in part 2.

Another consideration regarding trace morphing is that local behavior of traffic changes significantly during different stages. For instance, retrieving a web page has stages like TCP and HTTP connection establishment, sending request, receiving main page, receiving page dependencies, and downloading media contents. Since each stage of a trace has specific flow characteristics, morphing all packets of a trace to another trace may incur unnecessary bandwidth overhead and result in insufficient information leakage protection. We suggest a morphing approach in which source and destination traces are split into a fixed number of slices, $M$, and each slice of $T_{src}$ is morphed to the corresponding slice in $T_{dst}$. In evaluating the method, we consider different values for $M$, including $M=1$, that is the conventional morphing without slicing, from which $M=10$ was selected as the best candidate.

\subsubsection{Greedy Packet Morphing}

Greedy packet morphing is the algorithm, which operates on a user's device. This algorithm changes each traffic trace produced by online activities of user ($T_{src}$) on the fly minimizing the amount of information an adversary can obtain from observing the output traffic ($T_{out}$). It applies a greedy approach to determine how to change each packet size and timing, with the goal of producing a morphed trace that is similar to the selected $T_{dst}$, which is selected by the target site designator subsystem. The incoming and outgoing packets are interpreted and processed as two independent streams, only with the constraint of preserving the total timing order of packets among two streams. The greediness is due to the fact that for each packet, algorithm decides to either morph the packet or leave it unmodified regardless of future packets and solely based on two values of the goal function. Initially, $T_{dst}$ is divided into $M$ slices each containing $1/M$ of total packets. Packets of $T_{src}$ are also grouped into $M$ slices. Then, packets of each source slice are morphed to be as similar as the packets of corresponding destination slice with a reasonable overhead measured by the goal function and SimThresh parameter. In this step, each source packet is either changed to match size and timing of a packet in target slice, or is passed unmodified. Matching target trace packets improves similarity and passing reduces overhead, and the decision for each packet is made using a goal function establishing a trade-off between DiP and overhead. The goal function is discussed in part 2.

\begin{algorithm}[t]
\SetAlgoNoLine
\SetKwInOut{Input}{input}\SetKwInOut{Output}{output}

\Input{$T_{src}$ as an stream of packets, $T_{dst}$}
\Output{$T_{out}$: $T_{src}$ morphed in a way that will deceive adversaries to detect it as $T_{dst}$}
$S \leftarrow$ Divide $T_{dst}$ into $M$ slices\;
$s \leftarrow 0$\;
$i_{dst} \leftarrow 0$\;
\For{each packet P in $T_{src}$}{
\eIf{$GF_{morph}(pass) > GF_{morph}(morph)$}{
SEND $P$ unmodified;
}{
SEND $packetMorph(P, S[s][i_{dst}])$\;
$i_{dst} \leftarrow i_{dst} + 1$\;
}
\If{$moveToNewSlice(T_{src})$}{
$s \leftarrow s + 1$\;
$i_{dst} \leftarrow 0$\;
}
\If{$Ov>Tr_{Ov} \text{ and } GF_{similarity} < SimThres$}{
Send packets of current slice in $T_{src}$ unmodified\;
$s \leftarrow s + 1$\;
$i_{dst} \leftarrow 0$\;
}
}
\caption{Trace Morphing in TG-PSM}
\label{TraceMorphingAlg}
\end{algorithm}

\subsubsection{Overhead Controlling Criteria} A major concern in using traffic anonymization PETs are the bandwidth overhead and sending delays, that interfere with normal usage of users. A high overhead may result in user abandoning the use of PET completely (which results in zero protection against adversaries). Thus ensuring a overhead limit for PETs will result in more widespread use and improved total protection, even if such constraints may reduce information hiding performance. In our proposed method, we considered overhead implications in all design decisions and used techniques to limit the overhead incurred on resulting traffic. In particular, in determining how to morph each packet and in when a trace slice is similar enough to the target slice, we incorporated goal functions (overhead controlling criteria) that take into account the overhead of each action.

As shown in Algorithm \ref{TraceMorphingAlg}, each packet in $T_{src}$ is either morphed or passed unmodified. This is decided by calculating $GF_{morph}$ for both cases and only morphing the packet if the overhead introduced by morphing is justifiable with the resulting increase in goal function value. The goal function is defined as:
\begin{equation}
GF_{morph} = \displaystyle\sum_{i=0}^{k} dist(T_{out}[i], T_{dst}[i]) - \frac{\alpha}{D} \times \text{Overhead}.
\end{equation}
Where $dist$ can be any similarity measure, for which we used Jaccard coefficient, $D$ is the parameter specifying sensitivity of data (we used IM $\bmod{ \: nClust}$ as D), meaning the overhead tolerance by MU for that particular website, and $\alpha$ is a balancing factor among similarity and overhead. The summation is done over the part of traces already sent and the packets being processed, denoted by the variable $k$.

Additionally, for limiting the overhead, if the packets of one of the two active slices is finished, morphing is only continued until reaching a maximum similarity of output trace. This criteria is checked in lines 15-19 of Algorithm \ref{TraceMorphingAlg}, with the goal function defined as:
\begin{equation}
GF_{similarity} = dist(T_{out}, T_{src}).
\end{equation}
Both $dist$ function and $GF_{similarity}$ use Jaccard coefficient which in this context are defined as:
\begin{gather*}
\mathit{Generic \: Jaccard \: Coefficient:} \: \frac{A \cap B}{A \cup B} \\
\mathit{In \: TG-PSM:} \\
t_{i}: \: <site_{id_{i}},ps_{i}> \:,\: t_{j}: \: <site_{id_{j}},ps_{j}> \\
t_{i} \cap t_{j} = \{p \: | \: size(p) \in size(ps_{i}) \land size(p) \in size(ps_{j}) \} \\
t_{i} \cup t_{j} = \{p \: | \: size(p) \in size(ps_{i}) \vee size(p) \in size(ps_{j}) \}
\end{gather*}

\subsection{Finding Target Site and Tunability}
As a prerequisite for every website morphing algorithm, in addition to a $T_{src}$, which MU wants to hide, there should be a $T_{dst}$. In TG-PSM, unlike Wright's DTS and TM that choses $T_{dst}$ randomly, $T_{dst}$ is chosen according to the $T_{src}$ and the IM provided by the MU. Here, when the target site designator subsystem receives $<T_{src},IM>$, it choses the $IM \bmod{nClust}$ farthest cluster of websites and then randomly choses a website out of that cluster. The extracted website is known as $T_{dst}$. Then, $<T_{src},T_{dst},IM \bmod{nClust},SimThres>$ is passed to the traffic morphing subsystem. The algorithm is shown in algorithm \ref{FindingTargetSiteAlg}.
\newline

\begin{algorithm}[t]
\SetAlgoNoLine
\SetKwInOut{Input}{input}\SetKwInOut{Output}{output}
\Input{$\{<T_{src},IM>\}$}
\For{each T in \{$<T_{src},IM>\}$}
{
\eIf{IM == 0}{
$T_{dst} \leftarrow$ select a random instance from $C_{src}$}
{
$i \leftarrow C_{src}$\; 
$j \leftarrow CD[i,IM]$ \\ 
"Choosing $IMth$ farthest cluster" \\
$C_{dst} \leftarrow j$\;
$T_{dst} \leftarrow$ select a random instance from $C_{dst}$\;}
}
\Output{$\{T_{dst}\}$}

\caption{Finding Target Site in TG-PSM}
\label{FindingTargetSiteAlg}
\end{algorithm}

\subsubsection{Trace Representation}

In order to cluster different websites based on their behavior, we first needed a unified feature space for all websites. For this purpose, two contributing factors should be considered. First, how to exhibit a behavior of a website and second, because each website can have multiple traces, how to choose a single trace that best represents the website. Based on these two challenges, we present our two-step algorithm to represent a website. \par

\textbf{Step 1:} For each trace in our trace pool, we divide the trace into $M$ slices based on its packet count and calculate the mean size for incoming and outgoing packets. Hence each slice is presented in the form of $<Mean_{up},Mean_{down}>$. We repeat this process for all $M$ slices of each trace forming $SR_{ij}$ meaning feature vector for $jth$ trace belonging to $site_{id} = i$. \par
\textbf{Step 2:} For all traces of a website, we calculate the local outlier factor (LOF) \cite{breunig2000lof} using its 2M feature space (M slices, two measures from each). The trace with lowest LOF is considered the candidate trace. The reason behind using LOF is twofold. Firstly, because in the process of gathering traces timeouts, broken connections and network partitioning are possible. Hence, using mean or median instance is not suitable. Secondly, LOF uses only the local instances to determine the outliers. This attribute is crucial because a local outlier, might not appear to be an outlier when all of the instances are considered. Because of these two reasons, we believe LOF choses the best trace as a candidate trace for each website. \par

Another important feature of this representation scheme is that, this scheme demonstrates the behavior of any given trace in $M$ phases. Thus providing better accuracy and behavior matching. The algorithm of this scheme is presented in algorithm \ref{Trace Feature Space and Candid Selection}.

\begin{algorithm}[t]
\SetAlgoNoLine
\SetKwInOut{Input}{input}\SetKwInOut{Output}{output}
\Input{WS,TR,TF}

\For{each $t$ in $TR$}
{
Divide $t$ into $M$ slices based on $|t|$\;
\For{$s$ in $1:M$}
{
consider slice $sth$\;
Add $<Mean_{up},Mean_{down}>$ to $FeatSpace_{t}$\;
}
}
\For{each $ws$ in $WS$}
{
$ Tr \leftarrow \{ T | site_{id}(T) == ws\}$\;
\For{each $t$ in $Tr$}
{
$LOF_{t} \leftarrow LOF(t)$\;
}
$Can_{ws} \leftarrow $  trace with $min(LOF)$\;
Add $Can_{ws}$ to Cand\;
}
\Output{Cand}

\caption{Trace Feature Space and Candid Selection}
\label{Trace Feature Space and Candid Selection}
\end{algorithm}

\subsubsection{Clustering Instances}

In this section, the desirable attributes a practical clustering scheme in TG-PSM should possess are discussed.  In \cite{han2011data} Han et.al. introduced 11 attributes useful in cluster analysis, however, here we only discuss the ones we think are relevant to our framework .These attributes are: 
\begin{itemize}
\item \emph{High intra cluster similarity:} for creating indistinguishability between target websites when choosing from the same cluster, and also for predictive and controlled limit of overhead when morphing to a cluster.
\item \emph{Low inter cluster similarity:} for creating distinguishable target clusters, which in turn provides varied ranges and options for tunability.
\item \emph{Non-fuzzy membership:} for creating separated target clusters, which their overall estimation of the resulting overhead if morphed to, are predicted beforehand.
\item \emph{Resistant to noise:} because of the noisy nature of internet traffic, clustering in WF scenarios need to consider this and maintain resilience against such cases.
\item \emph{Scalability:} in order for the solution to be widely used and able to cope with high demands of morphing source and targets.
\end{itemize}

\begin{table}[t]
\centering\small
\begin{tabular*}{\linewidth}{@{\extracolsep{\fill}}p{0.2\linewidth}p{0.5\linewidth}p{0.3\linewidth}@{}}
\toprule
Parameter & Meaning & Value \\
\midrule
	cCount & Number of Clusters & 10 \\
	sCount & Number of Slices & 10 \\
	D & Tuning Factor & 1,3,5,7,9 \\
	Alpha & Bandwidth Impact Factor & 0.01 \\
	BW-D-Factor & Convert D to BW Threshold & 25 \\
	SimThresh & Min Similarity Two Slices & 0.7 \\
	cAlgorithm & Used Clustering Algorithm & PAM10 \\
	t & Training Set & 16 \\
	T & Testing Set & 4 \\
	K & Number of Websites to Morph & 32,128 \\
	N & Website Pool Size & 775 \\
         \bottomrule
\end{tabular*}
\newline
\caption{Used Parameters and Values in Evaluation}
\label{ParametersValues}
\end{table}

Based on candid traces extracted in previous part, we partition traces into mutually exclusive clusters. The used feature space in this step in a $2M$ dimension, all numeric vector for each candid trace. These $2M$ features are discussed in previous sections. We slice each trace into M parts and for each part calculate the mean-size of upstream and downstream packets and add them as features. In order to choose the best clustering algorithm and the best number of clusters to fully demonstrate dynamic tunability, we cluster our Candid set of traces using 5 different algorithms. These five algorithms~\cite{han2011data} are model-based clustering (MBC), partitioning around medoids (PAM), self organizing maps (SOM), agglomerative nesting (AGNES) and divisive hierarchical Clustering (DIANA). In all these algorithm, number of clusters are between 8 to 10. This decision is due to the fact that we need to have enough cluster in order to show the impact of dynamic tunability. After this step, we calculate 9 different indexes for all 15 clustering outcomes \cite{brock2011clvalid,desgraupes2013clustercrit}. Out of this 9, 4 are stability measures, showing the quality of cluster if one of the features are removed. The remaining 5 are internal indexes, These indexes examine the compactness of each cluster in addition to separability of them.The result of this examinations can be seen in appendix1. \\ Finally, based on these measures and the 5 attributes mentioned above and the calculated indexes, we choose PAM algorithm with 10 clusters as the best scheme to demonstrate dynamic tunability. The results in section VI all use PAM10 as default clustering. The values of all 9 metrics for 15 algorithms are presented in the Appendix \\

\section{Evaluation}

\subsection{Evaluation Methodology}

For all countermeasures except TG-PSM in accordance with \cite{dyer2012peek} , we randomly select $k$ websites out Herrmann dataset. For each of the $k$ websites, we extract two subset of traces belonging to the same website: $t,T$, t as the training set and T as the test set. For our trials, we set $t$ and $T$ to be 4 and 16, respectively, where there are least 20 traces for that website. In cases that 20 exceeds the total number of traces, 2 and 4 were used. The countermeasure is applied to each trace in both subsets and then they were given to the classifier. The accuracy is defined as ($\frac{c}{Tk}$) where c is the number of correctly guessed traces. Each trial is repeated 10 times and the result is the average accuracy. The only difference while evaluating TG-PSM is that, because websites are clustered, $t$ is chosen from $C_{src}$ while $T$ is chosen from the $C_{dst}$. These evaluation methodology is compatible with the work of Dyer et.al.~\cite{dyer2012peek} and is the literature standard for calculating the effectiveness of WF attacks and defenses. 

\begin{table*}[tbp]
\centering
\begin{tabular}{p{0.15\linewidth}p{0.35\linewidth}p{0.15\linewidth}p{0.15\linewidth}}
\hline
Countermeasure & Parameter & overhead & Classifier Accuracy \\
\hline
Traffic Morphing & $SimThresh = 0.7$ & 69.8 & $25.78\pm0.9$ \\
BuFLO & $\tau=0, \rho=40, d=1000$ & $96.6$ & $20.7$ \\
BuFLO & $\tau=0, \rho=20, d=1500$ & $147.0$ & $15.3$ \\
BuFLO & $\tau=10000, \rho=40, d=1000$ & $183.7$ & $16.2$ \\
BuFLO & $\tau=10000, \rho=20, d=1000$ & $347.2$ & $8.1$ \\
TAMARAW & $L_1$ & $84.3$ & $22.2\pm0.7$ \\
TAMARAW & $L_2$ & $251.0$ & $4.9\pm0.3$ \\
TGPSM & $D=1$ & $90.3$ & $14.63\pm0.8$ \\
TGPSM & $D=3$ & $122.7$ & $13.90\pm1.1$ \\
TGPSM & $D=5$ & $133.1$ & $10.60\pm0.4$ \\
TGPSM & $D=7$ & $146.6$ & $9.2\pm0.6$ \\
TGPSM & $D=9$ & $155.2$ & $7.6\pm1.5$ \\
TGPSM & $D=9,\:s \alpha = 1$ & $206.9$ & $4.0\pm1.0$ \\
\hline
\end{tabular} 
\newline
\caption{Overhead and accuracy results for different countermeasures against LL classifier at k=128 on Herrmann dataset}
\label{LLK128Results}

\centering
\begin{tabular}{p{0.15\linewidth}p{0.35\linewidth}p{0.15\linewidth}p{0.15\linewidth}}
\hline
Countermeasure & Parameter & overhead & Classifier Accuracy \\
\hline
Traffic Morphing & SimThresh = 0.7 & 58 & $5.71\pm1.8$ \\
BuFLO & $\tau=0, \rho=40, d=1000$ & $95.7$ & $3.4\pm2.1$ \\
BuFLO & $\tau=0, \rho=20, d=1500$ & $137$ & $5.3\pm0.2$ \\
BuFLO & $\tau=10000, \rho=40, d=1000$ & $131.6$ & $4.8\pm0.3$ \\
BuFLO & $\tau=10000, \rho=20, d=1000$ & $293$ & $1\pm0.8$ \\
TAMARAW & $L_1$ & $85.2$ & $5.7\pm0.3$ \\
TAMARAW & $L_2$ & $254.4$ & $1\pm0.8$ \\
TGPSM & $D=1$ & $91.5$ & $4.3\pm0.8$ \\
TGPSM & $D=3$ & $123.2$ & $3.2\pm0.8$ \\
TGPSM & $D=5$ & $134.1$ & $1\pm0.8$ \\
TGPSM & $D=7$ & $146.3$ & $1\pm0.8$ \\
TGPSM & $D=9$ & $156.9$ & $1\pm0.8$ \\
TGPSM & $D=9,\: \alpha = 1$ & $211.6$ & $1\pm0.8$ \\
\hline
\end{tabular} 
\newline
\caption{Overhead and accuracy results for different countermeasures against HA classifier at k=128 on Herrmann dataset}
\label{HAK128Results}

\centering
\begin{tabular}{p{0.15\linewidth}p{0.35\linewidth}p{0.15\linewidth}p{0.15\linewidth}}
\hline
Countermeasure & Parameter & overhead & Classifier Accuracy \\
\hline
Traffic Morphing & SimThresh = 0.7 & 41 & $88.8\pm0.8$ \\
BuFLO & $\tau=0, \rho=40, d=1000$ & $94.8$ & $29.3\pm1.9$ \\
BuFLO & $\tau=0, \rho=20, d=1500$ & $137$ & $23.9\pm2.4$ \\
BuFLO & $\tau=10000, \rho=40, d=1000$ & $131.5$ & $12.1\pm2.1$ \\
BuFLO & $\tau=10000, \rho=20, d=1000$ & $323$ & $8.3\pm0.4$ \\
TAMARAW & $L_1$ & $86$ & $25.6\pm1.1$ \\
TAMARAW & $L_2$ & $252.9$ & $9.6\pm1.5$ \\
TGPSM & $D=1$ & $86.7$ & $18.2\pm0.3$ \\
TGPSM & $D=3$ & $126.3$ & $23.5\pm1.3$ \\
TGPSM & $D=5$ & $131.7$ & $21.2\pm0.6$ \\
TGPSM & $D=7$ & $145.1$ & $19.2\pm1.1$ \\
TGPSM & $D=9$ & $169.3$ & $15.8\pm0.9$ \\
TGPSM & $D=9,\: \alpha = 1$ & $209.49$ & $9.7\pm0.8$ \\
\hline
\end{tabular} 
\newline
\caption{Overhead and accuracy results for different countermeasures against PA classifier at k=128 on Herrmann dataset}
\label{PAK128Results}
\end{table*}

\subsection{Experimental Setup}

\begin{figure}[t]
\centering
\includegraphics[width=9cm, height=5cm]{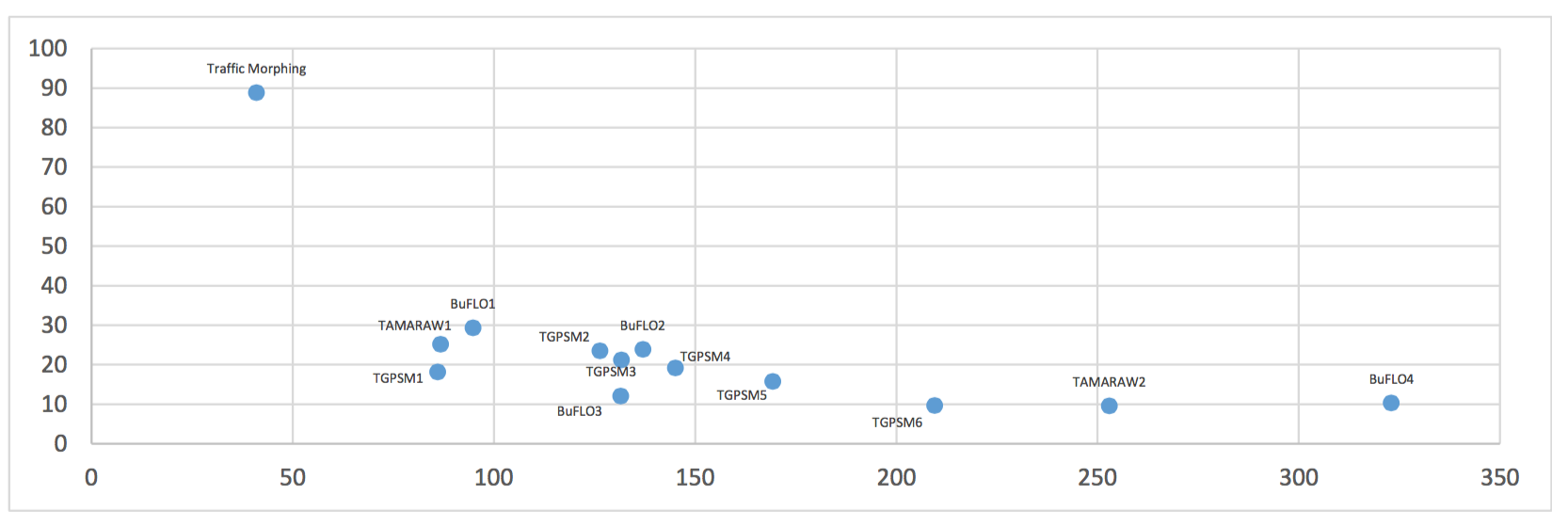}
\caption{DiP provided by different countermeasures with respect to bandwidth overhead incurred, errors of all data points are less than 5\% in each axis. Exact details in Table \ref{LLK128Results}}
\label{fig:ev1}
\end{figure}

For evaluating our proposed method, TG-PSM, we have incorporated our method into the evaluation framework provided by Dyer et.al. \cite{dyer2012peek}. The final framework consists of following components:

\begin{itemize}
 \item \emph{Evaluation Routine} this performs the sampling, training, and testing of classifiers and countermeasures and calculates accuracy of the classifier and bandwidth and timing overhead for the countermeasure being tested.
 \item \emph{Data Retrieval} this provides an abstract layer for uniform handling of different datasets and storage mechanisms, in addition to a in-memory caching layer for improving performance.
 \item \emph{Reference Implementations} this contains a reference implementation for a number of notable countermeasures and classifiers.
\end{itemize}
Use of this framework enables us to test TG-PSM using multiple existing datasets and attacks (traffic classifiers). Additionally, it enabled us to compare the results to existing countermeasures in an standard setup used by similar researches. To fully integrate our method that contains clustering and learning steps into this framework, we needed to make some improvements to it. We implemented TG-PSM and added it to the reference implementation component of framework, introduced some refactoring and redesigns, and slightly modified the evaluation routine to incorporate extra needed steps and testing methods. Additionally, to compare the results with similar cluster-based researches on the same dataset, also we imported existing implementations or implemented the countermeasures proposed in those researches. To verify our version of implementations, we conducted experiments similar to original researches and compared the results of our implementation to the original reported results, which were convincingly similar. Also, a test suite was developed for each newly implemented countermeasure to verify the expected behavior for some crafted traces. Then we used the verified implementations to run extra experiments with varying universe size and classifier-countermeasure combination and used the results to compare with the results obtained for TG-PSM. The results reported in next section are based on evaluating countermeasures in discussed framework using the following parameters: (List and values of all used parameters are shown in Table \ref{ParametersValues}.)

\begin{figure} [t]
\centering
\includegraphics[width=9cm, height=5cm]{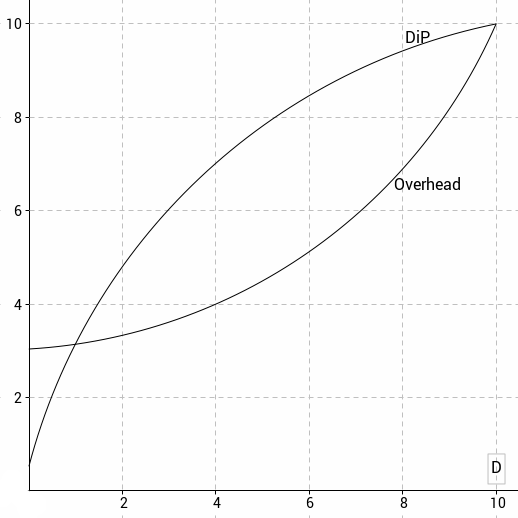}
\caption{Effect of changing D on DiP and overhead, reported as the percentage of maximum observed value for each metric. The values of D are discrete and the lines are only drawn for better interpretation.}
\label{fig:ev2}
\end{figure}

\begin{itemize}
 \item \emph{k=32/128}. The traces used for training and testing are all selected from these 2 different $K$-websites, forming the privacy set. The size of privacy set used in our setup, k, is 32/128, respectively.
 \item \emph{dataset=Herrmann}. Main dataset used for selecting traces is the Herrmann dataset \cite{herrmann2009website}, which is more recent than other commonly used dataset provided by Liberatore \cite{liberatore2006inferring} and is likely to have less invalid traces \cite{dyer2012peek}.
 \item \emph{classifiers=HA,LL,PA}. The three classification algorithms used are those introduced by Herrmann et. al. (HA), Liberatore and Levine (LL) \cite{liberatore2006inferring}, and Panchenko et. al. (PA) \cite{panchenko2011website}. The LL classifier provides a relatively fast algorithm with acceptable accuracy rates, and the PA classifier that usually performs best regarding to accuracy among existing classifiers, however, is slower to run for large traces and values of k \cite{dyer2012peek}.
\end{itemize}

\subsection{Results}

In this section we provide a summary of the results obtained from evaluating various countermeasures in the setup explained in the previous section. Two of the main concerns regarding PETs are DiP and the incurred overhead, which are usually related. In figure \ref{fig:ev1}, each point represents a traffic analysis countermeasure with parameters set to specific values, as defined in Table III. Points with lower accuracy values are suitable for use in cases where the traffic is sensitive and high overhead can be tolerated to achieve higher levels of security, whereas points with lower overhead values are more likely to be used by casual users with modest anonymity needs. In the bottom right section it can be observed that allowing a level of overhead, TG-PSM can provide protection comparable to algorithms with proven security like TAMARAW ($T_2$) \cite{cai2014systematic} and BuFLO ($B_4$). With lowering the overhead tolerance, TG-PSM starts to outperform similar countermeasures with same accuracy-overhead characteristics. Also, it is noticeable that TG-PSM can cover a wide range of MU security and performance requirements using only simple modifications to the tune parameter $D$.

A key element in TG-PSM is the easy tunability of the algorithm by MU, which boosts its usability for users. The effect of tuning parameter D in the resulting DiP and overhead is shown in figure \ref{fig:ev2}. The two curves corresponds to DiP and overhead metrics, and the percentage for each value of $D=D_0$ is the ratio of metric for $TGPSM(D=D_0)$ to the highest value of that metric. Generally, as D increases, DiP improves and as a result the accuracy of traffic classification is decreases, with a proportional increase in bandwidth overhead. This predictable behavior of TG-PSM with altering D can help users to tune the algorithm easily using only the parameter $D$ to achieve the level of protection required for each browsing session. The five instances of TG-PSM shown in figure \ref{fig:ev1} as $TGPSM_1$, $TGPSM_2$, $TGPSM_3$, $TGPSM_4$, $TGPSM_5$, $TGPSM_6$ correspond to values $1$, $3$, $5$, $7$, $9$ and $(9, \: \alpha=1)$ for D.

As an extra observation, we have separately measured the overhead of applying TG-PSM to traces of different total size in terms of aggregate packet length for different classes of trace lengths. We observe that for the traces with packet count of less than 1000 packets and total size of less than 10MB, the overhead is roughly $15\pm4\%$ less in average, and up to $25\pm4\%$ for traces with fewer than 500 packets. Consequently, we expect the real-word values of overhead imposed by TG-PSM on traffic traces to be noticeably less than the values reported above using the trace datasets, due to the fact that in common browsing experience traces are more probable to be of smaller sizes. More detailed results of our evaluation is presented in the Appendix.

\subsection{Comparison to Similar Algorithms}

The TG-PSM algorithm has two family of similar algorithms, one consisting of morphing-based algorithms and the other consisting of clustering-based countermeasures. Due to the use of clustering for selecting morphing target and use of overhead controlling criteria, TG-PSM outperforms the algorithms of first family as well as algorithms designed to protect against ideal attackers \cite{cai2014systematic}, in terms of DiP, overhead, or both. This comparison was done in more details in evaluation results in the previous section. Glove countermeasure \cite{nithyanand2014glove}, based on the similar idea of clustering websites for more intelligent adding of cover traffic to original traces, can represent the second family of algorithms. By comparing \cite{nithyanand2014glove} and \cite{cai2014systematic}, we concluded that against PA classifier, TAMARAW performs better and enjoys higher DiP. For this reason, we compare TG=PSM with TAMARAW empirically and only discuss the difference between TG-PSM and Glove. TG-PSM differs from Glove in following main aspects:

\textbf{Fine-grained tunability:} Using the tune parameter and the overhead controlling criteria, users can precisely choose the level of protection they need for each group of websites or each browsing session. Also, it helps users to impose a level of maximum tolerance for overhead based on available resources. The tunability can be done dynamically by the user for each use and does not require pre-computation of clusters. \par
\textbf{Traffic sectioning:} Due to the differences of traffic behavior in different phases of network protocols, for example the process of retrieving a webpage and its resources or handshakes and initial data transfers common in protocols, we have used a technique based on sectioning the traffic traces into multiple phases. This sectioning technique creates distinct morphing results that improves the efficiency of TG-PSM. \par
\textbf{Morphing-based approach:} Glove uses the clustering as a basis for adding cover traffic to achieve comparable results with lower bandwidth overhead compared to algorithms like BuFLO. This approach can be seen as a kind of smart alteration of countermeasures like PadMTU that focus on covering the important aspects of traffic. In contrast, TG-PSM aims to mislead the traffic analysis efforts by changing the original traces to traces similar to other real traffic, selected in a way that provides the most confusion in traffic classification. 

\section{Discussion}

\subsection{Open-World and Close-World}

In the course of this work, all trials are in close-world environment. In close-world environment, the attacker has a list of interesting websites and would like to know which one MU is visited. Also, the MU does not visit any website outside this list. In open-world environments on the other hand, MU can visit sites outside the interested list and attacker would like to know if any of interesting websites are visited. We believe because close-world environments favor the attacker, the countermeasures that we evaluate, will improve in open-world environments; hence, however effective countermeasures such as TG-PSM are in our evaluation, their open-world performance will be much better. In 2014 Juarez et. al. \cite{Juarez2014} showed the difference between classifier accuracy in these two environments and argued that in multi tab scenarios the accuracy of Dyer's N-grams classifier can be reduced by a drastic margin.

\subsection{Website Behavior in Different Phases of Trace}

As was proven by Panchenko's classifier, packet sequence behavior features are able to leak a great amount of information from traffic. They are also the reason for the failure of most of previous countermeasures. We believe a successful countermeasure should be able to detect and hide this class of features. As a requirement for this, a good representation of website behavior in different phases of loading and responding is required. In this paper, we used a 20-fold feature space that inspected this behavior in 10 phases each represented by 2 metric.

\subsection{Misleading Rather Than Confusing}

With regard to the countermeasure effect on classifier, there are two main approaches: Confusing and Misleading. In confusing approach, the classifier doesn't detect the right website and produces false prediction. Misleading on the other hand, fools the classifier into thinking another website was visited. The second approach (which is used by morphing countermeasure family such as TG-PSM) is specially useful in situation with white-list policy enforcements, where only a list of predetermined applications are allowed and the rest are banned.

\subsection{Correct Methodology for Analyzing Defense Capabilities}

The common methodology for evaluating the DiP of WF attacks is that classifier is both trained on $t$ and $T$ while both are countermeasure applied. This methodology suggests that either attacker knows the countermeasure mechanism or gathers the $T_{o}$ in MU's environment. We believe this assumption is not always right and is one of the reasons for poor performance of WF countermeasures. Although for compatibility with the literature in this field, we used the common methodology, in our opinion the right way must be that classifier is trained on website traffic without the counter measure ($t$) and is tested with the countermeasure applied traffic ($T$).

\section{Conclusion}

In this paper, we introduce a website fingerprinting countermeasure referred to as TG-PSM, which is designed to mislead the attacker into detecting false websites as the actual visited website of a user. TG-PSM morphs the traffic while considering the importance of the particular instance and the overall affordable bandwidth for that instance by the user. While previous countermeasures have proven to be inadequate and/or overly costly, we designed and implemented TG-PSM with the following attributes: \par
\textbf{Morphing rather than uniformization:} Uniformization countermeasures, although proven effective, suffer from high enforced bandwidth overhead and timing delay. Also, they fail to hide the proportional size of the trace, which can be used by packet sequence features. In TG-PSM, we use morphing approach that attempts to to mislead the attacker into detecting the wrong visited website. \par
\textbf{Dynamic tunability:} TG-PSM tunes its DiP and enforced bandwidth overhead bases on real-time user preferences. This feature is due to the fact that in the initial morphing process, TG-PSM chooses a target site which it believes will match the affordable overhead and desired DiP by the user. \par
\textbf{On-the-fly greedy stream morphing:} While previous morphing countermeasures needed the packet size distribution of source website, TG-PSM morphs the packet on the fly using a greedy algorithm which considers the created DiP and the resulting bandwidth overhead. This attribute results in more widespread use and better coverage for different website and if used, various applications. \par
Our evaluation results show that TG-PSM performs better than the previous morphing and uniformization website fingerprinting countermeasures both in the DiP and enforced bandwidth, e.g. providing better DiP while mainainig bandwidth overhead for D=1 and Providing better overhead while maintaining the same 
DiP for D=9 in k=128 situations.

\section{Acknowledgement}

For the results to be reproducible, our modifications to the base framework by Dyer et.al. is publicly available as a fork of the original repository\footnote{https://ce.sharif.edu/$\sim$ffani/TG-PSM}. This repository includes reference implementation for TG-PSM as well as the testing routine used for evaluation, and the code for clustering websites.

\bibliographystyle{ieeetr}

%


\end{document}